\documentclass[aps,prl,twocolumn,superscriptaddress,showpacs]{revtex4}
\usepackage{graphicx}
\usepackage{amsmath,amsfonts,amssymb}

\newcommand{\mtimes}{{$\times$}}

\newcommand{\rsrt}{{$\sqrt{7}\times\sqrt{3}$}}

\begin{document}

\title{Macroscopic Superconducting Current through a Silicon Surface Reconstruction with Indium Adatoms: Si(111)-(\rsrt)-In} 
\author{Takashi Uchihashi}
\email{UCHIHASHI.Takashi@nims.go.jp}
\affiliation{International Center for Materials Nanoarchitectonics, National Institute for Materials Science, 1-1 Namiki, Tsukuba 305-0044, Japan}

\author{Puneet Mishra}
\affiliation{International Center for Materials Nanoarchitectonics, National Institute for Materials Science, 1-1 Namiki, Tsukuba 305-0044, Japan}

\author{Masakazu Aono}
\affiliation{International Center for Materials Nanoarchitectonics, National Institute for Materials Science, 1-1 Namiki, Tsukuba 305-0044, Japan}

\author{Tomonobu Nakayama}
\affiliation{International Center for Materials Nanoarchitectonics, National Institute for Materials Science, 1-1 Namiki, Tsukuba 305-0044, Japan}

\date{\today}

\begin{abstract}
Macroscopic and robust supercurrents are observed by direct electron transport measurements on a silicon surface reconstruction with In adatoms (Si(111)-(\rsrt)-In).
The superconducting transition manifests itself as an emergence of the zero resistance state below 2.8 K.
$I-V$ characteristics exhibit sharp and hysteretic switching between superconducting and normal states with well-defined critical and retrapping currents.
The two-dimensional (2D) critical current density $J_\mathrm{2D,c}$ is estimated to be as high as $1.8 \ \mathrm{A/m}$ at 1.8 K.
The temperature dependence of $J_\mathrm{2D,c}$ indicates that the surface atomic steps play the role of strongly coupled Josephson junctions.

\end{abstract}


\maketitle

The state-of-the-art nanotechnology has enabled fabrication of ultrathin superconductors of high crystallinity and with atomically controlled thicknesses and interfaces \cite{Guo_PbQW,Eom_PbFilm,Oezer_PbHardSuper,Oezer_PbAlloy,Nishio_PbIsland,Qin_Pb2ML,Brun_PbFilm,Bollinger_SILaSrCuO}.
This has opened ways to tune superconductivity \cite{Guo_PbQW,Oezer_PbHardSuper,Bollinger_SILaSrCuO} and to investigate the thinnest crystalline layers for its emergence \cite{Eom_PbFilm,Qin_Pb2ML,Brun_PbFilm}.
The scope of the research on superconducting films has been substantially widened compared to the conventional studies where the samples were limited to granular and amorphous films \cite{Goldman_2DSuper,Katsumoto_2DSuper}.

Notably, superconductivity was found to exist for silicon surface reconstructions with metal adatoms \cite{Zhang_PbIn1ML}, which are the ultimate forms of thin epitaxial films.
This is of primary importance because a variety of metal-adsorbed semiconductor surfaces \cite{Lifshits_SurfacePhases} should now be regarded as candidates for new superconducting materials.
The above finding is, however, based on spectroscopic evidence of superconducting energy gaps observed by scanning tunneling microscopy (STM).
Surface supercurrents could be very fragile, but such information was not available. 
The local superconductivity shown there does not even guarantee the presence of macroscopic supercurrent because a surface is inevitably severed by numerous atomic steps.
The presence of macroscopic supercurrent can be best shown by electron transport measurements.
Such a measurement has, however, been hampered by the requirements of an ultrahigh vacuum (UHV) and low-temperature environment and of stable contacts to surface atomic layers.
The need to separate the surface current flow from the bulk contribution is also crucial \cite{YooWeitering_Si100,Uchihashi_1DIn}.

In this Letter, we perform direct and macroscopic electron transport measurements on a silicon surface reconstruction with In adatoms (Si(111)-(\rsrt)-In) \cite{Zhang_PbIn1ML,Kraft_R7R3,Rotenberg_R7R3,Yamazaki_R7R3} in UHV at low temperatures.
The superconducting transition is evidenced by observations of the zero resistance state and of $I-V$ characteristics exhibiting sharp and hysteretic switching below 2.8 K.
This macroscopic supercurrent also shows a significant robustness; the two-dimensional (2D) critical current density $J_\mathrm{2D,c}$ is estimated to be as high as $1.8 \ \mathrm{A/m}$ at 1.8 K.
The observed temperature dependence of $J_\mathrm{2D,c}$ indicates that the surface atomic steps serve as strongly coupled Josephson junctions.

Figure 1(a) shows an atomic structural model of Si(111)-(\rsrt)-In surface (referred to as (\rsrt)-In) proposed by Kraft \textit{et al.} \cite{Kraft_R7R3}, where one In atom corresponds to one unit cell of the Si(111)-1\mtimes1 surface.
This surface hosts two-dimensional nearly free electrons with a high density \cite{Rotenberg_R7R3}, exhibiting metallic transport properties down to low temperatures \cite{Yamazaki_R7R3}.
The central parallelogram indicates the \rsrt\ unit cell.
Note that the precise positions of In atoms have not been determined and are assumed to be the hollow sites of the Si(111)-1\mtimes1 here.
The surface was prepared by thermal evaporation of In onto a clean Si(111) substrate followed by a brief annealing in UHV \cite{Kraft_R7R3,Rotenberg_R7R3,Yamazaki_R7R3}, and was subsequently characterized by low energy electron diffraction (LEED) and STM.
Sharp LEED spots (Fig. 1(b)) and high-resolution STM images (Fig. 1(c)) confirmed the presence of a well-ordered (\rsrt)-In phase \cite{footnote_R7R3phases}.
For the large-scale morphology, STM observations show that the surface consists of flat terraces separated by atomic steps with a height of 0.31 nm (see Fig. 1(d)(e)).
The averaged terrace width was estimated to be 370 nm from repeated measurements.

\begin{figure}
\includegraphics[width=80mm]{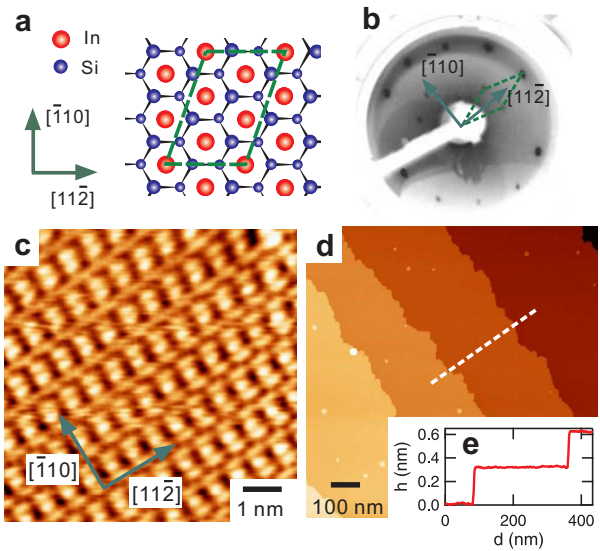}
\caption{(Color)
(a) Atomic structural model of the (\rsrt)-In surface proposed by Kraft \textit{et al.} \cite{Kraft_R7R3}. The red and blue spheres represent In and surface Si atoms, respectively. The dashed lines indicate the (\rsrt) unit cell. The In atoms are assumed to be on the hollow sites of the Si(111)-1\mtimes1 for simplicity. (b) LEED pattern taken at beam energy $E = 86$ eV. The dashed lines indicate the reciprocal unit cell. (c) High-resolution STM image taken at a sample voltage $V_s = -0.015$ V and a tunneling current $I_t = 270$ pA. (d) Large-scale STM image taken at $V_s = 2 $ V and $I_t = 160$ pA. (e) Height profile taken along the dashed line in (d).
}
\label{Fig1}
\end{figure}

To perform four-terminal resistance measurements, the sample surface was patterned as depicted in Fig. 2(a).
The central square with a size of $1\times1 \ \mathrm{mm^2}$ and the four outer squares connected at corners (purple areas) are made of the (\rsrt)-In surface.
The latter serves as current and voltage terminals for the former, thus enabling van der Pauw's measurements \cite{vanderPauw,Tegenkamp_1DPb}.
The rest of the sample surface (gray areas) consists of bare Si, which is prepared by $\mathrm{Ar^{+}}$ sputtering through a shadow mask (see Supplemental Materials (SM) 1) \cite{Supplemental}.
Following the sample preparation, four Au-coated spring probes were brought into contact with the current/voltage terminal patterns in a UHV-compatible cryostat.
Four-terminal zero bias resistance $R_0$ and $I-V$ characteristics of the (\rsrt)-In surface were measured from 20 to 1.8 K.
Here $R_0$ is simply defined as the ratio of the measured voltage to the bias current.
The sample was carefully shielded from room temperature radiation and Si-diode thermometers were thermally anchored to prevent an erroneous reading.
To access the electron conduction only through the (\rsrt)-In surface at low temperatures, Si(111) substrates without intentional doping (resistivity $\rho > 1000 \ \mathrm{\Omega cm}$) were used \cite{Uchihashi_1DIn}.
The whole procedure including the pattern fabrication and the transport measurement was performed without breaking the UHV environment \cite{Uchihashi_1DIn,Uchihashi_Nanostencil}.

\begin{figure}
\includegraphics[width=70mm]{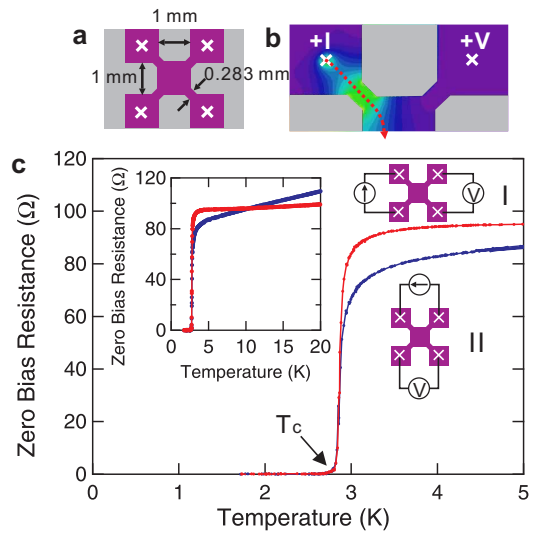}
\caption{(Color)
(a) Drawing of the sample patterning. The central and the four outer squares (purple areas) are made of (\rsrt)-In, and the surrounding regions (gray areas) consist of sputtered Si surfaces.
(b) Calculated current density distribution on the sample. The bright green (dark purple) represents a high (low) current density. The red dotted line indicates the flow of a bias current.
(c) Temperature dependence of zero bias resistances. $R_\mathrm{0,I}$ (red lines) and $R_\mathrm{0,II}$  (blue lines) are zero bias resistances measured using the probe configurations I and II, respectively, with DC bias currents of 1 $\mathrm{\mu A}$. Left inset: $R_\mathrm{0,I}$ and $R_\mathrm{0,II}$ for a larger temperature range. Right insets: schematic drawings of the probe configurations I and II.
}
\label{Fig2}
\end{figure}

Figure 2(c) exemplifies the temperature dependence of zero bias resistance.
$R_\mathrm{0,I}$ and $R_\mathrm{0,II}$ were measured with two complimentary configurations I, II by rotating the pairs of current/voltage probes by 90$^\circ$  (see the right insets of Fig. 2(c)).
DC bias currents of 1 $\mathrm{\mu A}$ were supplied and the offset voltages due to the thermoelectric effect were removed by inverting the bias polarity.
Followed by gradual decreases with decreasing temperature $T$, $R_\mathrm{0,I}$ and $R_\mathrm{0,II}$ dropped to nearly zero simultaneously below $T = 2.8$ K, indicating a superconducting transition.
The residual resistances for $T < 2.6$ K are negligibly small compared to the noise level of 0.2 $\Omega$.
The fact that both $R_\mathrm{0,I}$ and $R_\mathrm{0,II}$ became zero excludes a possibility of failure to detect a voltage drop due to an extremely high transport anisotropy.
Although the transition is sharp just above the onset of the zero resistance state (ZRS), it exhibits a precursor below $\sim 4$ K which is absent for pure bulk superconductors.
This is attributed to the superconducting fluctuation effects inherent to 2D superconductors \cite{Aslamasov_2DSuper}.
The transition temperature $T_c = 2.8$ K determined from the onset of the ZRS is slightly lower than the previously reported $T_c = 3.14$ K, which was determined from the opening of superconducting energy gap \cite{Zhang_PbIn1ML}.

Assuming a homogeneous and isotropic resistivity for simplicity, we calculated the current distribution within the present sample using the finite element method (Fig. 2(b)).
This allows us to determine the ratio $\rho_\mathrm{2D}/R_0$ to be 4.54, where $\rho_\mathrm{2D}$ is the sheet resistance (2D resistivity) and $R_0$ the four-terminal resistance defined as above.
If $(R_\mathrm{0,I}+R_\mathrm{0,II})/2$ is identified with $R_0$, $\rho_\mathrm{2D} \approx 410 \ \mathrm{\Omega}$ at 5 K is obtained.
This value is sufficiently smaller than the critical sheet resistance $h/4e^2 (= 6.45 \times 10^3 \ \mathrm{\Omega})$ of the superconductor-insulator transition, which is the criterion for the emergence of global superconducting coherence \cite{Goldman_2DSuper,Katsumoto_2DSuper}.
Let us mention that the surface has actually some anisotropy depending on samples and locations.
The origin of the anisotropy can be attributed to the local directions of surface steps, which should be the dominant electron scatterers at low temperatures.
 This may also cause the different temperature dependences of $R_\mathrm{0,I}$, $R_\mathrm{0,II}$ displayed in Fig. 2(c).

A further evidence for the presence of supercurrent was obtained by measuring $I-V$ characteristics.
The main panel of Fig. 3(a) shows a series of $I-V$ characteristics taken by sweeping DC bias currents at different temperatures from 3.11 K to 1.77 K.
The bias current was swept toward the increasing direction at a rate of 87 $\mathrm{\mu A/s}$ and the probe configuration I was adopted.
The results were nearly independent of the sweeping rate between 23 and 350 $\mathrm{\mu A/s}$.
Below 2.8 K, the sample first switched from the normal state to the ZRS (where $dV/dI = 0$) at the retrapping current $I_r$ and then from ZRS to the normal state at the critical current $I_c$.
The switching behaviors became remarkably pronounced as the temperature was lowered down to 1.77 K.
The ZRS can be safely assigned to the superconducting state because it is destroyed by an excessive current.
By inverting the sweeping direction, the hysteresis of $I-V$ characteristics was confirmed as shown in the inset.
The origin of the hysteretic switching can be a Joule heating effect \cite{Skocpolt_HeatEffect}.
Figure 3(b) summarizes the temperature dependence of critical current $I_c$ (green squares) and retrapping current $I_r$ (pink squares).
The data taken with configurations I and II are shown by closed and open squares, respectively.
$I_c$ and $I_r$ were almost identical for the two configurations.
Following the onset around 2.8 K, both $I_c$ and $I_r$ steadily increase as temperature is lowered, reaching 520 $\mathrm{\mu A}$ and 230 $\mathrm{\mu A}$ at 1.8 K, respectively.

\begin{figure}
\includegraphics[width=68mm]{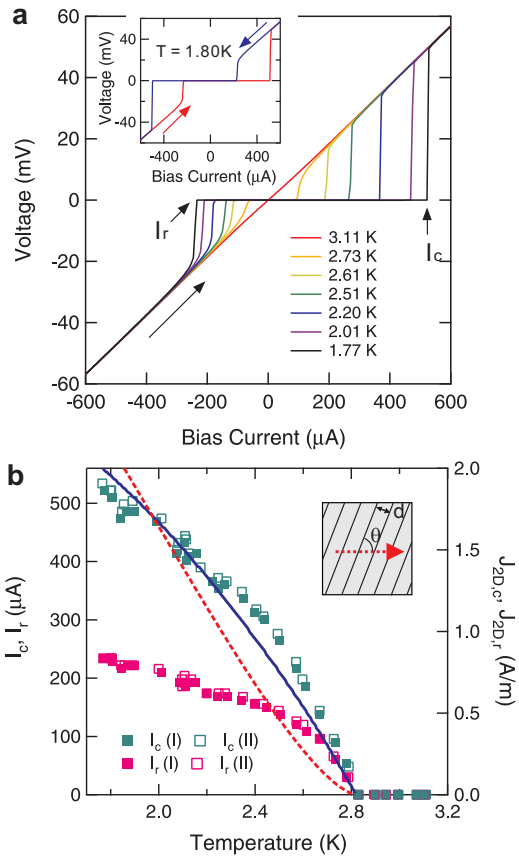}
\caption{(Color)
(a) Temperature dependence of $I-V$ characteristics obtained with configuration I.  The inset shows $I-V$ characteristics taken by inverting sweeping directions at 1.80 K. The arrows indicate the sweeping directions.
(b) Temperature dependences of critical current $I_c$ (green squares) and retrapping current $I_r$ (pink squares). The data taken with configurations I and II are shown by closed and open squares, respectively. The blue solid and red dotted lines show theoretical fits. For details, see the text.  The inset sketches the relation between current flow and atomic steps.
}
\label{Fig3}
\end{figure}

The 2D critical and retrapping current densities $J_{2D,c}$, $J_{2D,r}$ are determined as described below \cite{Supplemental}.
First we note that the self-screening of magnetic field produced by supercurrent is very weak because the superconducting layer is atomically thin in the present system (SM 2).
In addition, if the magnitude of the superconducting order parameter is constant, the distribution of supercurrent is identical to that of normal current when the same boundary condition is imposed (SM 3).
Therefore, the above calculation on the normal current density is also valid here.
Figure 2(b) shows that the current density is the highest at the constrictions between the current probes and the central area, which means that $I_c$ is determined in the constrictions (SM 4).
Since the current density is nearly constant in the middle of the constriction, the measured $I_c$ and $I_r$ can be converted to $J_{2D,c}$ and $J_{2D,r}$ by dividing them by its width $w_c=0.283 \ \mathrm{mm}$ (see the right axis of Fig. 3(b)).
$J_{2D,c}= 1.8  \ \mathrm{A/m}$ at 1.8 K is remarkably high considering that the conducting layer is single-atom thick.
If the thickness of (\rsrt)-In is assumed to be double the covalent radius of In (= 0.30 nm), this corresponds to a 3D critical current density $J_{3D,c} = 6.1 \times 10^9 \mathrm{A/m^2}$.

The mechanism of determining the critical current is discussed as follows.
At low temperatures, the atomic steps are considered to be the dominant source of elastic electron scattering and resistance on the surface.
In the superconducting state, therefore, they can serve as Josephson junctions.
In this case, the temperature dependence of $J_{2D,c}$ is given by the following equation \cite{Ambegaokar-Baratoff}:
\begin{equation}
J_{2D,c}(T)=\frac{\pi\Delta(T)}{2e\rho_{\mathrm{step}}}\tanh(\Delta(T)/2k_B T)  \label{eq:AB_equation},
\end{equation}
where $\Delta(T)$ is the superconducting energy gap at $T$, $\rho_{\mathrm{step}}$ the ($T$-independent) normal resistance of the atomic step for unit length, $k_B$ the Boltzmann constant.
Since $\Delta(T)$ follows the BCS theory \cite{Zhang_PbIn1ML} and $T_c=2.8$ K is already known, the experimentally obtained $J_{2D,c}(T)$ can be fitted with Eq. (\ref{eq:AB_equation}) using $J_{2D,c}(0)$ as the only parameter.
The fitting can reproduce the experiment semi-quantitatively, which gives $J_{2D,c}(0)= 2.7 \ \mathrm{A/m}$ (the blue solid line in Fig. 3(b)).
Another requirement of $\Delta(0)=0.57 \ \mathrm{meV}$ \cite{Zhang_PbIn1ML} allows us to determine $\rho_{\mathrm{step}}$ to be $3.3\times 10^{-4} \ \mathrm{\Omega m}$.
If the critical current is determined by Cooper pair breaking induced by an excessive supercurrent, it follows the equation $J_{2D,c}(T) \propto (1-(T/T_c)^2)^{3/2}$ \cite{pair_breaking}.
Fitting with the equation gives a poor result (red dotted line in Fig. 3(c)), excluding this mechanism.
This suggests that the supercurrent density is not high enough to substantially suppress the magnitude of order parameter \cite{Tinkham_Textbook}, in consistent with the earlier assumption on its spatial uniformity.
We note that the surface steps should be regarded as strongly coupled junctions because of the high critical currents, although the term Josephson junction is conventionally used for a weak coupling.

The normal resistance of atomic steps $\rho_\mathrm{step}$ obtained above can be compared to that from the normal sheet resistance $\rho_\mathrm{2D}$.
Suppose a local current flows across atomic steps at an angle of $\theta$ and the steps are separated by an average distance $d$ (see the inset of Fig. 3(b)).
If we simply assume that $\theta$ is randomly distributed between -$\pi/2$ and $\pi/2$ over the sample surface, the average sheet resistance $\rho_\mathrm{2D}$ can be calculated as
\begin{equation}
\rho_\mathrm{2D}=\frac{1}{\pi}\int_{-\pi/2}^{\pi/2} \frac{\rho_\mathrm{step}|\sin\theta|}{d} d\theta = \frac{2\rho_\mathrm{step}}{{\pi}d}. \label{eq:sheet_resistance}
\end{equation}
Insertion of experimentally obtained values $d=370 \ \mathrm{nm}$ and $\rho_\mathrm{2D}=410 \ \mathrm{\Omega}$ into Eq. (\ref{eq:sheet_resistance}) gives $\rho_\mathrm{step}=2.4 \times 10^{-4}\ \mathrm{\Omega m}$.
This is in satisfactory agreement with the value  $\rho_\mathrm{step}=3.3 \times 10^{-4}\ \mathrm{\Omega m}$ determined earlier, supporting the Josephson junction model of atomic steps. 
We note that, in the analysis of the Josephson junction, the angle $\theta$ between the current flow and steps was not taken into account.
This is because the critical current is determined by individual atomic steps (presumably by one with the highest $\rho_\mathrm{step}$) and the step separation $d$ is not relevant.
On the contrary, in the case of sheet resistance, a smaller $\theta$ increases the effective step separation as $d/\sin\theta$, resulting in a lower $\rho_\mathrm{2D}$.
We also note that the obtained step resistances are comparable to the previously reported $\rho_\mathrm{step}\approx 2\times 10^{-4}\ \mathrm{\Omega m}$ for Si(111)-($\sqrt{3}\times\sqrt{3}$)-Ag \cite{Matsuda_StepR}, although a different surface reconstruction was studied there.

In conclusion, we have demonstrated that macroscopic and robust supercurrents can run on a (\rsrt)-In surface despite the presence of atomic steps.
It was indicated that the surface atomic steps serve as strongly coupled Josephson junctions.
The present study makes various surface reconstructions of silicon and related semiconductors candidates for practical superconducting materials.
We envision that atomic-scale design and tuning of superconductivity will be feasible for such surface systems based on the current nanotechnology.

This work was financially supported by JSPS under KAKENHI Grant No. 21510110. T. U. thanks S. Hasegawa and Y. Ootuka for helpful discussions and Y. Wakayama for technical supports. U. Ramsperger is acknowledged for critical reading of the manuscript.



\begin{references}


\bibitem{Guo_PbQW}
Y. Guo \textit{et al.}, Science \textbf{306}, 1915 (2004).

\bibitem{Eom_PbFilm}
D. Eom, S. Qin, M.-Y. Chou, and C. K. Shih, Phys. Rev. Lett. \textbf{96}, 027005 (2006).

\bibitem{Oezer_PbHardSuper}
M. M. {\"O}zer, J. R. Thompson, and H. H. Weitering, Nature Phys. \textbf{2}, 173 (2006).

\bibitem{Oezer_PbAlloy}
M. M. \"Ozer, Y. Jia, Z. Zhang, J. R. Thompson, and H. H. Weitering, Science \textbf{316}, 1594 (2007).

\bibitem{Nishio_PbIsland}
T. Nishio \textit{et al.}, Phys. Rev. Lett. \textbf{101}, 167001 (2008).

\bibitem{Qin_Pb2ML}
S. Qin, J. Kim, Q. Niu, and C.-K. Shih, Science \textbf{324}, 1314 (2009).

\bibitem{Brun_PbFilm}
C. Brun \textit{et al.}, Phys. Rev. Lett. \textbf{102}, 207002 (2009).

\bibitem{Bollinger_SILaSrCuO}
A. T. Bollinger \textit{et al.}, Nature \textbf{472}, 458 (2011).

\bibitem{Goldman_2DSuper}
A. M. Goldman and  N. Markovic, Phys. Today \textbf{51}, 39 (1998).

\bibitem{Katsumoto_2DSuper}
S. Katsumoto, J. Low. Temp. Phys. \textbf{98}, 287 (1995).

\bibitem{Zhang_PbIn1ML}
T. Zhang \textit{et al.}, Nature Phys. \textbf{6}, 104 (2010).

\bibitem{Lifshits_SurfacePhases}
V. G. Lifshits, A. A. Saranin, and A. V. Zotov, \textit{Surface Phases on Silicon: Preparation, Structures, and Properties} (Wiley, Chichester, 1994).

\bibitem{YooWeitering_Si100}
K. Yoo and H. H. Weitering, Phys. Rev. Lett. \textbf{87}, 026802 (2001)

\bibitem{Uchihashi_1DIn}
T. Uchihashi and U. Ramsperger, Appl. Phys. Lett. \textbf{80}, 4169 (2002).

\bibitem{Kraft_R7R3}
J. Kraft, S. L. Surnev, and F. P. Netzer, Surf. Sci. \textbf{340}, 36 (1995).

\bibitem{Rotenberg_R7R3}
E. Rotenberg \textit{et al.}, Phys. Rev. Lett. \textbf{91}, 246404 (2003).

\bibitem{Yamazaki_R7R3}
S. Yamazaki \textit{et al.}, Phys. Rev. Lett. \textbf{106}, 116802 (2011). 

\bibitem{footnote_R7R3phases}
It is known that (\rsrt)-In has two forms of atomic configurations which are close in energy: (\rsrt)-In-hex and (\rsrt)-In-rect \cite{Kraft_R7R3}. Our repeated STM measurements indicate that, for our samples, (\rsrt)-In-hex was the dominant phase although (\rsrt)-In-rect coexisted.

\bibitem{vanderPauw}
L. J. van der Pauw, Philip. Res. Reps. \textbf{13}, 1 (1958).

\bibitem{Tegenkamp_1DPb}
C. Tegenkamp \textit{et al.}, Phys. Rev. Lett. \textbf{95}, 176804 (2005).

\bibitem{Supplemental}
See Supplemental Materials (EPAPS Document No.***) for the sample patterning (SM 1) and the analysis on spatial distribution of supercurrent (SM 2-4). 


\bibitem{Uchihashi_Nanostencil}
T. Uchihashi, U. Ramsperger, T. Nakayama, and M. Aono, Jpn. J. Appl. Phys. \textbf{47}, 1797 (2008).

\bibitem{Aslamasov_2DSuper}
L. G. Aslamasov and A. I. Larkin, Phys. Lett. \textbf{26A}, 238 (1968).


\bibitem{Skocpolt_HeatEffect}
W. J. Skocpolt, M. R. Beasley, and M. Tinkham, J. Appl. Phys. \textbf{45}, 4054 (1974).

\bibitem{Ambegaokar-Baratoff}
V. Ambegaokar and A. Baratoff,  Phys. Rev. Lett. \textbf{10}, 486 (1963); erratum, \textbf{11}, 104 (1963).

\bibitem{pair_breaking}
J. Bardeen, Rev. Mod. Phys. \textbf{34}, 667 (1962).


\bibitem{Tinkham_Textbook}
M. Tinkham, \textit{Introduction to Superconductivity}, 2nd Ed., Sect. 4.4 (McGraw-Hill Co., New York, 1996).


\bibitem{Matsuda_StepR}
I. Matsuda \textit{et al.}, Phys. Rev. Lett. \textbf{93}, 236801 (2004).







\end{references}
\end{document}